\documentclass[
 reprint,
 superscriptaddress,
 nolongbibliography,
 amsmath,amssymb,
 aps,
]{revtex4-2}

\usepackage{graphicx}
\usepackage{dcolumn}
\usepackage{bm}
\usepackage{booktabs}
\usepackage{threeparttable}
\usepackage{multirow}
\usepackage{gensymb}
\usepackage{siunitx}
\usepackage{xcolor}
\usepackage{array} 

\newcommand{\exciting}{{\usefont{T1}{lmtt}{b}{n}exciting}}

\newcommand{\eg}{{\it e.g.}, }
\newcommand{\ie}{{\it i.e.}, }

\begin{document}
\title{Novel Approach to Structural Relaxation of Materials in Optically Excited States}

\author{Mao Yang}
\email[]{maoyang@physik.hu-berlin.de}
\affiliation{Physics Department and IRIS Adlershof, Humboldt-Universit\"at zu Berlin, 12489 Berlin, Germany}
\affiliation{Max Planck Institute for Dynamics of Complex Technical Systems, Standtorstraße 1, 39106 Magdeburg, Germany}

\author{Claudia Draxl}
\affiliation{Physics Department and IRIS Adlershof, Humboldt-Universit\"at zu Berlin, 12489 Berlin, Germany}
\affiliation{European Theoretical Spectroscopy Facility (ETSF)}

\date{\today}

\begin{abstract}
We present a first-principles method for relaxing a material's geometry in an optically excited state. This method, based on the Bethe-Salpeter equation, consists of solving coupled equations for exciton wavefunctions and atomic displacements. Our approach allows for structural relaxation of excited states to be achieved through a single iteration. As results, one obtains not only energy and wavefunction of the thus modified, \ie self-trapped, exciton, but also the mechanism of relaxation in terms of atomic displacements in the respective phonon eigenmodes. We demonstrate and evaluate our formalism with the example of the three molecules CO, H$_{2}$O, and NH$_{3}$. 
\end{abstract}


\maketitle
Excitons, being created upon light absorption in a material or molecule, are electron-hole pairs bound by Coulomb interaction~\cite{Frenkel1931PRB}. Their generation implies a rearrangement of the electron density that can cause also the atomic configuration to rearrange. This, in turn, gives rise to a modified exciton, \ie altered excitonic wavefunction and energy. This new excited state dressed by atomic displacements is called a self-trapped exciton (STE)~\cite{Yu2010book,William1990JPCS}. 

STEs have a variety of interesting properties, thus attracting growing attention in condensed-matter physics and materials science, both experimentally and theoretically~\cite{li2019jpcl,Mack2019ACSPhoto,VanGinhoven2003JCP,Ismail2005PRL,Li2020FO,guo2022am,Benin2018ACIE,yin2020CM,mackrodt2022JCP}. For example, STEs can generate a broad luminescence spectrum covering the entire visible light range~\cite{Worku2020APLM,Zhiyuan2019APL,Mack2019ACSPhoto}. Halide perovskites, for instance, can achieve efficient and stable single-source white-light emission~\cite{Luo2018vj,Chen2021Acsnano}. Moreover, a massive Stokes shift of up to 8 eV caused by STEs has been demonstrated in alkaline-earth-metal fluorides and alkali halides~\cite{Pooley1970JPC,William1990JPCS}. This value, being one to two orders of magnitude larger than a typical phonon energy, implies that lattice vibrations cannot be treated as simple perturbations with respect to excitons.

The state-of-the-art approach to compute electronic and optical excitations from first principles, is many-body perturbation theory (MBPT)~\cite{hedin1965PR,Onida2002RMP}. The Bethe-Salpeter equation (BSE)~\cite{strinati1988} typically carried out on top of a ground-state calculation based on density-functional theory (DFT)~\cite{kohn1996JPC,Kohn1965PR} followed by the $GW$~\cite{hybertsen1986prb} approximation for obtaining the electronic quasi-particle energies, has been very successful in describing optical spectra of semiconductors, including exciton wavefunctions and binding energies for a variety of materials~\cite{vorwerk2018JPCL,Gonzalez2022PRM,Rohlfing2000PRB,qiu2013PRL,Sander2015PRB,MolinaSanchez2018ACSPEM,Palummo2020AEL}. The interplay between excitons and lattice distortions (vibrations), particularly the structural relaxation of STEs, remains, however, a significant challenge for first-principles calculations.

An intuitive way to perform a structural relaxation in the excited state is via minimization of atomic forces analogous to ground-state calculations. An analytic form of excited-state forces was first derived in the field of quantum chemistry ~\cite{stanton1995JCP}. The unfavorable scaling with (at least) the 6th power in system size makes it, however, a very expensive method. Another excited-state force approach, with 4th-power scaling, was based on the Bethe-Salpeter Hamiltonian~\cite{Ismail2005PRL}. In this approach as well as in constrained DFT (CDFT), as often used for core excitations (the so-called supercell core-hole approach~\cite{Olovsson2009PRB}), a large supercell capable of hosting an electron-hole pair is necessary to avoid spurious interaction of adjacent periodic replica in crystalline materials. Also finite systems, correspondingly, require a large simulation box. Overall, no matter which excited-state force method is adopted, it is hampered by the formidable computational cost. This is in particular so for crystalline materials where excitons may be delocalized over many unit cells.  

Other theoretical investigations of STEs are based on time-dependent density-functional theory (TDDFT)~\cite{Runge1984PRL,Kretz2021JCTC}. TDDFT is most popular for single molecules~\cite{Cocchi2015PRB}, but less reliable for electron-hole correlations and optical absorption spectra of crystalline materials when employing (semi)local exchange-correlation (xc) kernels, \eg the adiabatic local-density approximation (ALDA)~\cite{Onida2002RMP}.
 
In this work, we show how to overcome the limitations of above described methods. In order to accomplish this task, we apply the variational principle to the total energy of a system containing one exciton, using the exciton wavefunction and the atomic displacements as variables. This way, we establish equations for these quantities, coupled by the exciton-phonon (ex-ph) interaction matrix. Our formalism requires the exciton energies and wavefunctions from momentum-dependent BSE, the phonon normal modes and frequencies, and electron-phonon (e-ph) matrix elements from density-functional perturbation theory (DFPT)~\cite{Baroni2001RMP} or the frozen-phonon approach~\cite{Yin1980PRL}. From the solution of the coupled equations, we also obtain the STE energy as well as the atomic displacements corresponding to it. In addition, this method reveals inherently to what extent a particular phonon mode contributes to the structural relaxation in the excited state. As a first application of this novel method, we demonstrate results for three molecules (CO, H$_2$O, and NH$_3$), and compare them to those of the excited-state forces method as well as experiment.

We start by considering a system containing one exciton. The total energy of the system $E_\text{tot}$, a functional of the exciton wavefunction $\Psi(\mathbf{r}_\text{e}, \mathbf{r}_\text{h})$ and the atomic coordinates $\{ \mathbf{\tau}_{\kappa \alpha} \}$ of atom $\kappa$ along the Cartesian direction $\alpha$, is given by 
\begin{equation}\label{eq:tot_energy}
    \begin{aligned}
        E_{\text{tot}}&[\Psi(\mathbf{r}_\text{e},\mathbf{r}_\text{h}),\{\mathbf{\tau}_{\kappa \alpha}\}] \\
        &= E_{\text{G}}[\{\mathbf{\tau}_{\kappa \alpha}\}] + E_{\text{ex}}[\Psi(\mathbf{r}_\text{e},\mathbf{r}_\text{h}),\{\mathbf{\tau}_{\kappa \alpha}\}],
    \end{aligned}
\end{equation}
where $E_{\text{G}}$ is the ground-state energy. The exciton energy, $E_{\text{ex}}[\Psi(\mathbf{r}_\text{e},\mathbf{r}_\text{h}),\{\mathbf{\tau}_{\kappa \alpha}\}]$, 
 of the state with atomic configuration $\{ \mathbf{\tau}_{\kappa \alpha} \}$ can be obtained as the expectation value of the Bethe-Salpeter Hamiltonian that can be written for singlet excitations in the Tamm-Dancoff approximation as $\hat{H}^{\text{BSE}} = \hat{H}^{\text{diag}} + 2\hat{H}^{\text{x}} + \hat{H}^{\text{dir}}$. The diagonal term $\hat{H}^{\text{diag}}$ describes single-particle transitions without considering electron-hole correlations. $\hat{H}^{\text{x}}$ is the exchange term (missing for triplet excitations) and the direct term, $\hat{H}^{\text{dir}}$, is characterized by the screened Coulomb interaction. The BSE equation yields the exciton energies and wavefunctions by solving the corresponding eigenvalue problem,
\begin{equation}\label{eq:BSE}
    \sum_{v'c'} H^{\text{BSE}}_{vc,v'c'} \, \mathcal{A}^{\lambda}_{v'c'} = E^{\lambda} \mathcal{A}^{\lambda}_{vc}\,\, ,
\end{equation}
where $E^{\lambda}$ is the exciton energy with index $\lambda$. The exciton wavefunction is expressed as $\Psi_{\lambda}(\mathbf{r}_\text{e},\mathbf{r}_\text{h}) = \sum_{vc} \mathcal{A}^{\lambda}_{vc} \, \phi_{v}^{*}(\mathbf{r}_\text{h})  \, \phi_{c}(\mathbf{r}_\text{e})$, where $c$ and $v$ denote conduction and valence states, respectively, and $\mathbf{r}_\text{e}$ and $\mathbf{r}_\text{h}$ the corresponding electron and hole coordinates. $\mathcal{A}^{\lambda}_{vc}$ are the coefficients in the two-particle basis, chosen in practice as products of Kohn-Sham (KS) valence and conduction wavefunctions, $\phi_{v} (\mathbf{r})$ and $\phi_{c} (\mathbf{r})$, respectively. They convey information about the nature and composition of the exciton. Note that we drop the dependence on the {\bf k} vector in the Brillouin zone here as we will show results for molecules below; we emphasize though that the overall formalism is completely general.

To consider the interplay between the optically excited state and the atomic displacements, we expand $E_{\text{G}}$ in a Taylor series around the ground-state equilibrium positions $\{\mathbf{\tau}_{\kappa \alpha}^{0}\}$. Since the first-order derivatives of $E_{G}$ are equal to zero, the displacements, $\Delta \mathbf{\tau}_{\kappa \alpha} = \mathbf{\tau}_{\kappa \alpha} - \mathbf{\tau}_{\kappa \alpha}^{0}$, only contribute from the second-order term on. Truncating the series at second order, $E_\text{G}$ reads:
\begin{equation}\label{eq:G_expansion}
    \begin{aligned}
	E_\text{G} \lbrack \lbrace \mathbf{\tau}_{\kappa \alpha} \rbrace \rbrack
	= & \, E_\text{G} \lbrack \lbrace \mathbf{\tau}^{0}_{\kappa \alpha} \rbrace \rbrack \\
	 &+ \frac{1}{2} \sum_{\kappa \alpha,\kappa '\alpha '} \, F_{\kappa \alpha, \kappa ' \alpha '} \, \Delta \mathbf{\tau}_{\kappa \alpha} \, \Delta \mathbf{\tau}_{\kappa ' \alpha '}. 
    \end{aligned}
\end{equation}
Here, $F_{\kappa \alpha, \kappa ' \alpha '}$ are the force constants of the ground state. Performing a first-order Taylor expansion of the BSE Hamiltonian, we obtain:
\begin{equation}\label{eq:ex_expansion}
    \begin{aligned}
    &E_{\text{ex}}[\Psi(\mathbf{r}_\text{e},\mathbf{r}_\text{h}),\{\mathbf{\tau}_{\kappa \alpha}\}] = \int \! d \mathbf{r}_\text{e} d \mathbf{r}_\text{h} \, \times \\
    & \Psi^{*}(\mathbf{r}_\text{e},\mathbf{r}_\text{h}) \! \left[ \hat{H}_{0}^{\text{BSE}} + \sum_{\kappa \alpha} \frac{\partial \hat{H}_{0}^{\text{BSE}}}{\partial \mathbf{\tau}_
	{\kappa \alpha} } \Delta \mathbf{\tau}_{\kappa \alpha} \right] \! \Psi(\mathbf{r}_\text{e},\mathbf{r}_\text{h}).
	\end{aligned}
\end{equation}
Thereby, $\hat{H}_{0}^{\text{BSE}}$ indicates the BSE Hamiltonian in the ground-state geometry.

As the total energy, $E_{\text{tot}}$, of the STE in its equilibrium should be minimal, its functional derivatives with respect to the exciton wavefunction and the atomic displacements must vanish. Introducing the Lagrange multiplier $\epsilon$ to guarantee the normalization of the exciton wavefunction $\Psi(\mathbf{r}_\text{e},\mathbf{r}_\text{h})$,  the conditional-extremum equations can be written as
\begin{equation}
 	\frac{\delta}{\delta \Psi^*} \left[ E_{\text{tot}} - \epsilon \left( \! \int d \mathbf{r}_\text{e} d \mathbf{r}_\text{h}  \vert \Psi(\mathbf{r}_\text{e},\mathbf{r}_\text{h}) \vert ^2 - 1 \right) \right] \! = 0 \:,
 \end{equation}
 \begin{equation}
	\frac{\delta E_\text{tot}[\{\Psi(\mathbf{r}_\text{e},\mathbf{r}_\text{h})\},\{\mathbf{\tau}_{\kappa \alpha}\}]}{\delta \mathbf{\tau}_{\kappa \alpha}  }=0 \:,
\end{equation}
which finally lead to the following coupled nonlinear equations:
\begin{equation}\label{eq:STE_energy}
	\left[ \hat{H}^{0}_\text{BSE} + \sum_{\kappa \alpha} \frac{\partial \hat{H}_{0}^\text{BSE}}{\partial \mathbf{\tau}_
	{\kappa \alpha} } \, \Delta \mathbf{\tau}_{\kappa \alpha} \right] \! \Psi(\mathbf{r}_\text{e},\mathbf{r}_\text{h}) = \epsilon \, \Psi (\mathbf{r}_\text{e},\mathbf{r}_\text{h}) \:,
\end{equation}
\begin{equation}\label{eq:disp}
    \begin{aligned}
 	\Delta \mathbf{\tau}_{\kappa \alpha} = & - \! \sum_{\kappa ' \alpha '} F^{\ -1}_{\kappa \alpha,\kappa ' \alpha '} \, \times \\
  & \int \!\! d \mathbf{r}_\text{e} d \mathbf{r}_\text{h} \Psi^{*}(\mathbf{r}_\text{e},\mathbf{r}_\text{h}) \, 
  \frac{\partial \hat{H}_{0}^{\text{BSE}}}{\partial \mathbf{\tau}_{\kappa '\alpha '}} \,
  \Psi(\mathbf{r}_\text{e},\mathbf{r}_\text{h}) \, .
    \end{aligned}
 \end{equation}
From Eq.~\eqref{eq:STE_energy} we see that the Lagrange multiplier $\epsilon$ is exactly the exciton energy, $E_{\text{ex}}$, of the self-trapped state.

We now expand the STE wavefunction $\Psi(\mathbf{r}_\text{e},\mathbf{r}_\text{h})$ in terms of the eigenstates of the non-perturbed Hamiltonian $\hat{H}_{0}^{\text{BSE}}$, 
\begin{equation}\label{eq:exciton_expansion}
  \Psi(\mathbf{r}_\text{e},\mathbf{r}_\text{h}) = \sum_{\lambda} \mathcal{C}_{\lambda} \, \Psi_{\lambda}(\mathbf{r}_\text{e},\mathbf{r}_\text{h}),
\end{equation}
where the expansion coefficients $\{\mathcal{C}_{\lambda}\}$ are subject to normalization, \ie $\sum_{\lambda} \vert \mathcal{C}_{\lambda} \vert ^{2} = 1$. The atomic displacements in the excited state can be expanded in phonon eigenmodes of the unperturbed system~\cite{Sio2019PRB,Sio2019PRL}: 
\begin{equation}\label{eq:phonon_expansion}
  \Delta \mathbf{\tau}_{\kappa \alpha} = - 2 \sum_{\nu} \mathcal{B}_{\nu} \left(\frac{\hbar}{2 M_{\kappa } \omega_{\nu}} \right) ^ {\!\!{1}/{2}} \! \text{e}_{\kappa \alpha , \nu} \: ,
\end{equation}
where $M_{\kappa}$ is the mass of atom $\kappa$, $\omega_{\nu}$ the phonon frequency of the vibrational mode $\nu$, $\mathbf{e}_{\kappa \alpha, \nu}$ the $\alpha$ component of its eigenvector, and $\mathcal{B}_{\nu}$ the corresponding dimensionless amplitude. Substituting Eqs.~\eqref{eq:exciton_expansion} and~\eqref{eq:phonon_expansion} into Eqs.~\eqref{eq:STE_energy} and~\eqref{eq:disp}, we obtain the algebraic form of the coupled equations for $\mathcal{C}_{\lambda}$ and $\mathcal{B}_{\nu}$:
\begin{equation}\label{eq:algebraic_exciton}
    2 \sum_{\lambda', \nu} \mathcal{B}_{ \nu} \, \mathcal{G}^{*}_{\lambda \lambda' \nu} \, \mathcal{C}_{\lambda'} = (E^{\lambda} - \epsilon) \, \mathcal{C}_{\lambda},  
\end{equation}
\begin{equation}\label{eq:algebraic_disp}
    \mathcal{B}_{\nu} = \frac{1}{\hbar \omega_{\nu}}\sum_{\lambda \lambda'} \mathcal{C}^*_{\lambda} \, \mathcal{G}_{\lambda \lambda' \nu} \, \mathcal{C}_{\lambda'},
\end{equation}
where $\mathcal{G}_{\lambda \lambda' \nu}$ is the exciton-phonon matrix element between exciton states $\lambda$ and $\lambda'$ via the vibrational mode $\nu$~\cite{Chen2020PRL} which can be calculated as
\begin{equation}\label{eq:ex-el-phonon}
    \mathcal{G}_{\lambda \lambda' \nu} = \sum_{vcc'} \mathcal{A}^{\lambda *}_{vc} \, \mathcal{A}^{\lambda '}_{vc'} \, g_{cc'\nu} - \sum_{vv'c} \mathcal{A}^{\lambda *}_{vc} \, \mathcal{A}^{\lambda '}_{v'c} \, g_{v'v\nu} \, .
\end{equation}
The electron-phonon matrix elements, $g_{mn \nu}$, are defined as~\cite{Giustino2017RMP}
\begin{equation}\label{eq:eph_matrix}
    g_{mn \nu} = \sum_{\kappa \alpha}\left(\frac{\hbar}{2 M_{\kappa} \omega_{\nu}} \right) ^ {\!{1}/{2}} \!\! \text{e}_{\kappa \alpha , \nu} \! \int \!\! d \mathbf{r} \; \phi^{*}_{m} (\mathbf{r}) \, \frac{\partial V^0 _{\text{KS}} }{\partial \mathbf{\tau}_{\kappa \alpha} } \, \phi_{n} (\mathbf{r}),
\end{equation}
where $V_{\text{KS}}^{0}$ is the KS potential~\cite{Kohn1965PR} in the atomic ground-state configuration. Combining Eqs.~\eqref{eq:ex-el-phonon} and~\eqref{eq:eph_matrix}, one can determine $\mathcal{C}_{\lambda}$ and $\mathcal{B}_{\nu}$ by solving Eqs.~\eqref{eq:algebraic_exciton} and~\eqref{eq:algebraic_disp} iteratively. Substituting $\mathcal{C}_{\lambda}$ and $\mathcal{B}_{\nu}$ into Eqs.~\eqref{eq:exciton_expansion} and~\eqref{eq:phonon_expansion}, we finally obtain the STE wavefunction in real space and the atomic displacements in cartesian coordinates. 

Eqs.~\eqref{eq:algebraic_exciton}--\eqref{eq:eph_matrix} constitute the main achievement of this work. Notably, their solution only requires the exciton energies $E^{\lambda}$ and coupling coefficients $A^{\lambda}_{vc}$ from the BSE in the original primitive cell. Also the e-ph matrix elements $g_{mn \nu}$ can be calculated by linear response theory in the primitive cell.  This formalism is applicable not only to molecules but --recovering the dependence of all quantities on the crystal momentum-- also to periodic solids, including materials that exhibit delocalized excitons.

We now apply our method to the molecules CO, H$_{2}$O, and NH$_{3}$, for which both experimental and theoretical results of excited states are available in literature. To do so, we have implemented our formalism into the all-electron full-potential code \exciting~\cite{Gulans2014JCP} that employs the linearized augmented planewave + local orbital (LAPW+lo)  basis. We perform DFT calculations in the local-density approximation (LDA) for the ground-state in a 10$\times$10$\times$10~\AA$^3$ cubic cell that is large enough to avoid spurious interactions between molecular replica. Muffin-tin radii of 0.7, 0.9, 1.0, and 1.0 bohr are used for H, C, N, and O, respectively. For molecules, only $\Gamma$ point excitons need to be considered that are computed by solving the BSE. A basis-set cutoff of $R_{\text{MT}}G_{\text{\text{max}}} = 4$ is used in all calculations where $R_{\text{MT}}$ is the radius of the smallest muffin-tin sphere, and $G_{\text{max}}$ is the planewave cutoff. We carefully checked the number of occupied and unoccupied bands to ensure convergence of the BSE calculations. Phonons and e-ph matrix elements are computed by the frozen-phonon approach and also verified by DFPT.
\begin{table}[h]
    \caption{Calculated equilibrium geometries in terms of bond length $d$ (in \AA) and bond angle $\theta$ (in \degree) of selected molecules in their excited states compared to experiment. The symmetry of the excited state for CO, H$_2$O, and NH$_3$ is $1^{1}\Pi$, $2^{1}B_{1}$, and $1^{1}A_{2}$, respectively.}
    \label{tab:Str_Rel}
    \centering
    \setlength{\tabcolsep}{2.1mm}{
    \begin{threeparttable}
        \begin{tabular}{lccccc}
        \toprule
        \multirow{2}{*}{Method} & CO & \multicolumn{2}{c}{H$_{2}$O} & \multicolumn{2}{c}{NH$_{3}$} \cr
        \cmidrule(lr){2-2} \cmidrule(lr){3-4} \cmidrule(lr){5-6}
                           & $d$    & $d$  & $\theta$ & $d$  & $\theta$  \cr
        \midrule
        This work          & 1.22   & 0.99 & 108.5    & 1.07 & 119.7     \cr
        CDFT~\cite{Ismail2003PRL} & 1.21   &  -    & -         & 1.08 & 120.0       \cr
        BSE-ESF~\cite{Ismail2003PRL} & 1.26   & -    & -        & 1.08 & 120.0       \cr
        QCA~\cite{stanton1995JCP,Furche2002JCP,McCarthy1987JCP} & 1.21-1.22 & 0.96   & 109.0          & 1.06 & 120.0       \cr
        Experiment~\cite{Furche2002JCP}  & 1.24   & 1.02 & 107.0      & 1.08 & 120.0       \cr
        \bottomrule
        \end{tabular}
    \end{threeparttable} }
\end{table}
\begin{figure}[htb]
  \begin{center}
    \includegraphics[width=0.45\textwidth]{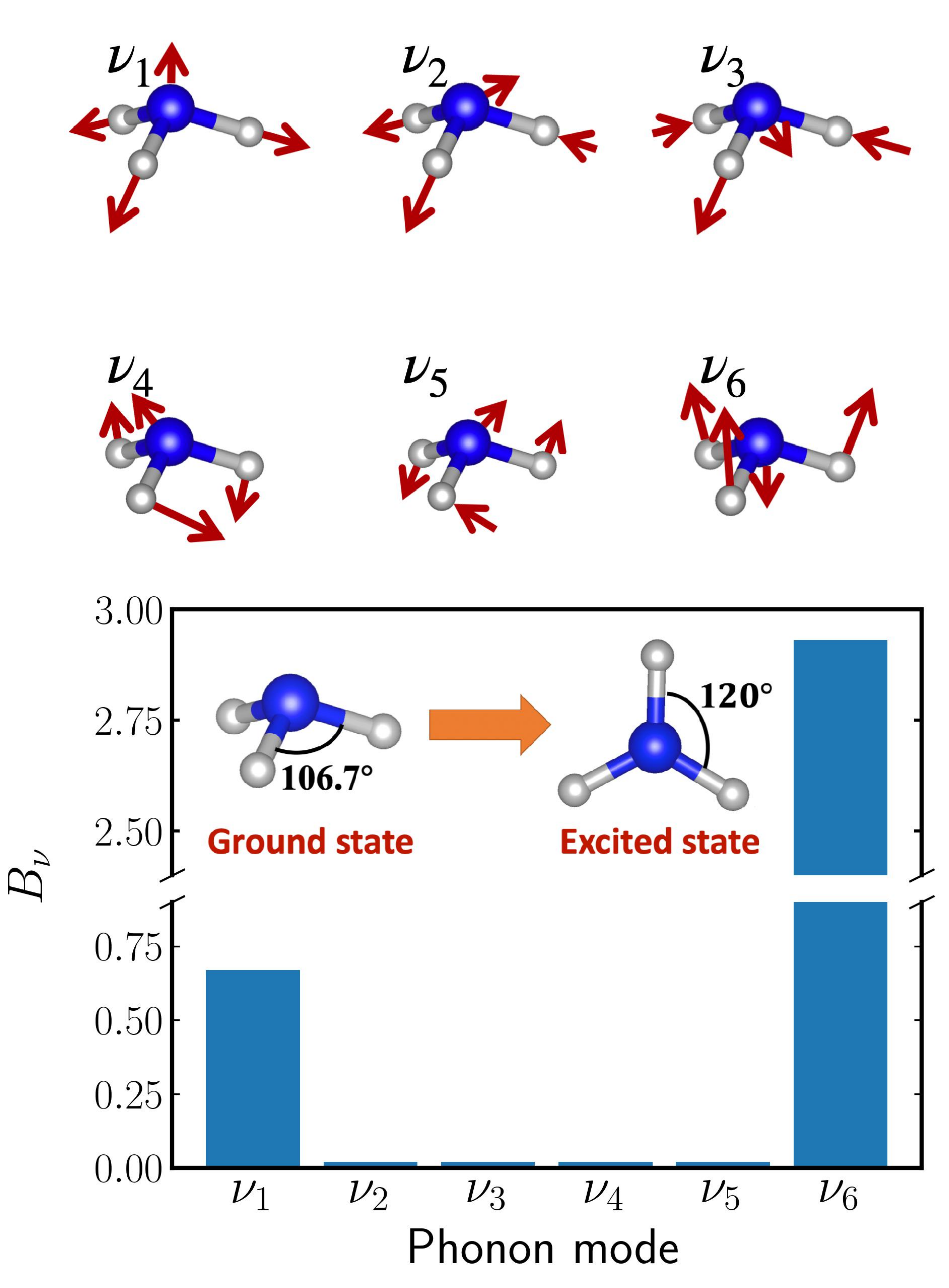} 
  \end{center}
    \caption{Amplitude of phonon modes $B_{\nu}$ contributing to the structural relaxation of the excited state of ammonia  (bottom panel). Blue and gray balls represent N and H atoms, respectively. The six phonon modes displayed in the top panel are symmetric-stretch ($\nu_{1}$), asymmetric-stretch ($\nu_{2}$, $\nu_{3}$), bending ($\nu_{4}$, $\nu_{5}$) and umbrella mode ($\nu_{6}$). The main contributions come from the symmetric stretch and the umbrella mode, while the others are negligible.}
    \label{fig:ammonia_ph}
\end{figure}

Our results for the excited-state geometries of CO, H$_2$O, and NH$_3$, in terms of bond lengths, $d$, and bond angles, $\theta$, are presented in Table~\ref{tab:Str_Rel}. They are in excellent agreement with results from three other computational methods, namely CDFT, the BSE based excited-state force (BSE-ESF) and the above mentioned quantum-chemistry approach (QCA). All of them also match experiment very well. Aanalyzing the exciton wavefunctions reveals them to be composed to a large extent (more than 93\%) of transitions between the highest occupied molecular orbital (HOMO) and the lowest unoccupied molecular orbital (LUMO) in all three cases. This explains that the LUMO and HOMO wavefunctions in CDFT calculations are able to capture the essential characteristics of the exciton wavefunctions and the molecules may therefore acquire the geometrical structure with similar precision as in the other approaches. 

Figure \ref{fig:ammonia_ph} illustrates the structural relaxation mechanism of the lowest singlet excited state for the example of the ammonia molecule (NH$_{3}$). The geometry undergoes a transition from a pyramidal ground-state structure with an H-N-H angle of 106.7\degree\ to a flat excited-state structure with a H-N-H bond of 120\degree. Based on the solution of Eq.~\eqref{eq:algebraic_disp}, we find that a substantial contribution with an amplitude of 0.67, comes from the mode $\nu_{1}$, which extends the N-H bond from 1.02 \AA\ to 1.07\AA. The umbrella mode, $\nu_{6}$, opening the H-N-H angle from 105\degree\ to 119.7\degree, is, however, the dominant contribution with an amplitude of 2.93. Since both modes are symmetric, the equilibrium structure in the excited state exhibits also high symmetry. The four asymmetric modes ($\nu_{2}$ -- $\nu_{5}$) contribute only very little.
 
In summary, we have developed a novel first-principles method that allows us to determine the structural relaxation of molecules as well as periodic systems in optically-excited states. Based on the Bethe-Salpeter equation of MBPT, we have derived coupled equations for atomic displacements and the exciton wavefunction to account for exciton-phonon interaction that drives the relaxation process. Moreover, the phonon spectral decomposition of the atomic configuration reveals the underlying microphysical mechanisms. In this work, we have provided examples for molecules that could be compared to available literature results, showing excellent agreement and demonstrating the capability of our approach. We emphasize, however, that the procedure is general enough to capture periodic materials that are out of reach for existing methods due to their computational costs. Taking advantage of phonon coordinates and momentum-dependent BSE, we can obtain all desired quantities considering only a primitive simulation cell rather than a supercell. The complete formalism for perodic systems will be published in a forthcoming manuscript. Enabling hence efficient calculations, our work, allows for addressing a variety of problems of optically excited states, among them self-trapped excitons, which are observed in many materials.

This work was supported by the German Research Foundation, project number 182087777 (SFB 951). We are grateful to Ignacio Gonzalez Oliva and Sebastian Tillack for providing us with code to calculate e-ph matrix elements prior to publication, and to Alex Buccheri for reviewing the code. We also thank Pasquale Pavone, Sebastian Tillack, and Manoar Hossain for their critical reading of the manuscript.
\bibliography{literature.bib}
\end{document}